\documentclass[prb,twocolumn]{revtex4}

\def\D{{\cal{D}}}
\def\E{\varepsilon}
\def\t{\tilde}
\def\e{{\cal{E}}}
\usepackage{graphicx}
\begin{document}
\title{Two-dimensional Hubbard-Holstein bipolaron}
\author{
A.\ Macridin$^{1,3}$,  G.\ A.\ Sawatzky $^{1,2}$ and Mark Jarrell$^{3}$}

\address{$^{1}$University of Groningen, Nijenborgh 4, 9747 AG, Groningen, The Netherlands \\
$^{2}$University of British Columbia, 6224 Agricultural Road, Vancouver, 
Canada\\
$^{3}$University of Cincinnati, Cincinnati, Ohio, 45221, USA }

\date{\today}

\begin{abstract}

We present   a diagrammatic Monte Carlo  study of  the properties of the 
Hubbard-Holstein bipolaron on a  two-dimensional square lattice.
With a small Coulomb repulsion, $U$,   and with increasing electron-phonon
interaction, and when  reaching a value about two times smaller than the one corresponding to 
the transition of light polaron to heavy polaron,
the system suffers a sharp transition from a state  formed by two 
weakly bound light polarons to a heavy, strongly bound  on-site  bipolaron.
Aside from this rather conventional bipolaron a new bipolaron state is found 
for large U at intermediate and large electron-phonon coupling, corresponding to two polarons 
bound on nearest-neighbor sites.  
We discuss both the properties of the different bipolaron states and  the 
transition from one state to another.
We present a phase diagram in  parameter 
space defined by the electron-phonon coupling and $U$. 
Our numerical method does not use any artificial approximation and can be easily modified to 
other bipolaron models with  longer range electron-phonon and/or electron-electron  interaction.

\end{abstract}

\maketitle
\section{Introduction}

\label{sec:intoduction}
The interaction between electrons and lattice degrees of freedom plays a crucial role in  the  
properties of many materials and results in a multitude of physical phenomena. Structural transitions like 
the cooperative Jahn-Teller distortion in perovskites, the Peierls dimerization in one-dimensional systems, 
or the pairing and
the condensation of the charge carriers  in superconductivity are some of the most spectacular effects
which originate from electron-lattice interaction. Polaronic and bipolaronic effects are found 
in many materials like transition metal-oxides~\cite{tmo}, superconducting materials~\cite{spm} 
and conjugated polymers~\cite{cjp}. 
In the last years there has been  growing experimental evidence that even in the fashionable strongly correlated
materials like manganates and cuprates,  aside from the unscreened Coulomb repulsion the electron-lattice 
interaction is  extremely important~\cite{mo,co}. 

The theoretical investigation of the interaction between  charge carriers and  lattice vibrations 
has
a long history. The concept of a polaron, which describes an electron which carries with it a lattice deformation,
was introduce first by Landau in 1933~\cite{landau}. One of the most successful theories of the last century, 
the Bardeen, Cooper and Schrieffer (BCS) theory of superconductivity~\cite{BCS} 
is based on the observation that phonons
induce an effective attraction between electrons.

The discovery of high $T_{c}$ superconductors renewed the interest in the study of the electron-phonon 
interaction, and the bipolaron problem in particular. In the cuprate materials the strength of the
interaction between  electrons and phonons is believed to be 
in the intermediate  regime. Because of this rather strong interaction the lattice 
ions change their equilibrium position when they are in the vicinity of charge carries, i.e 
the charge carries drive a phonon vacuum instability~\cite{alexandrov,alexandrov1}. 
The classical Migdal-Eliashberg approach to the theory of superconductivity~\cite{migdal},
which neglects these effects and is valid only for small electron-phonon coupling, cannot be applied to cuprates.
Therefore special theoretical attention was given to  scenarios where electrons (or holes)
form pairs in real space (bipolarons), which suffer Bose-Einstein condensation, leading to
superconductivity~\cite{alexandrov}. In this respect 
Alexandrov and co-workers proposed the strong electron-phonon coupling theory
as a starting point for explaining the physics of high $T_c$ superconductors.
In their theory~\cite{alexandrov,alexandrov2}, as a consequence of the strong electron-phonon
interaction, the holes pair in bipolarons. For low charge carriers density, the system can be regarded as 
a charged $2e$ Bose gas which condense at $T_c$, resulting in superconductivity.

However the study of these systems is complicated by the failure of both strong and 
weak coupling perturbation theory, even for simple model systems, at 
intermediate coupling strength. Novel algorithms were
developed to address the problem in this difficult region of parameters space.
The bipolaron problem, which is defined by two electrons on a lattice,
has been intensively studied in the last years.
For the one dimensional case the problem was addressed by Bon$\tilde{c}$a 
{\em et al.}~\cite{bonca} with  an  exact diagonalization technique on a variationally  
determined Hilbert subspace. 
Other one-dimensional calculations were based on
variational methods~\cite{magna} and  density-matrix combined with Lancszos diagonalization 
technique~\cite{fehske}.
The two-dimensional case was investigated in the adiabatic approximation~\cite{aubry} and with 
variational methods~\cite{iadonisi}.

In this paper we address the  Hubbard-Holstein (HH) bipolaron which is 
one of the  simplest and 
most popular models which contains both electron-electron 
and the electron-phonon interaction.
Its solution is important for understanding 
the competition between the phonon-induced electron-electron  attraction and the 
electron-electron Coulomb repulsion\footnote{However the reader should be aware that  
HH model   is presumably not the best 
candidate to address the physics of cuprate superconductors, at intermediate and large electron-phonon coupling
producing bipolarons with extremely large effective mass~\cite{chak}. Models with longer range  
electron-phonon interaction~\cite{bonca1} are believed to be more realistic for describing cuprates.}. 
As our calculation, based on a  Quantum Monte Carlo algorithm, shows, in the intermediate and 
strong electron-phonon coupling regions, phonons strongly renormalize the effective
hopping integral of the electrons, strongly reduce the effective on-site Coulomb repulsion but
do not significantly affect the nearest-neighbor exchange interaction. 
This gives rise to low-energy effective Hamiltonians with a 
large antiferromagnetic interaction relative to the
the effective hopping and the effective repulsion terms, which couldn't be derived  starting 
from pure electronic models.    
We find that, depending on the value of
the Coulomb repulsion, two electrons can form an on-site strongly bound
state for small  $U$ or  a weakly bound nearest-neighbor localized state
for larger $U$. The former state appears when the effective on-site attraction
due to  phonons overcome the Coulomb repulsion and the later is a result of
the exchange interaction which wins over the strongly renormalized electron (polaron)
kinetic energy.

We developed a Diagrammatic Quantum Monte Carlo (DQMC) algorithm
suitable for studying the two-dimensional HH bipolaron.
To our knowledge, this is the first 
two-dimensional calculation which considers
dynamical phonons and does not entail any artificial truncation of the Hilbert space. 
Our algorithm computes the imaginary time two-particle Green's function from which we
extract information about the bipolaron state  at
long imaginary time.
The DQMC algorithms were introduced by Prokof'ev {\em et al.}~\cite{prokofev}
and used to calculate the properties of Fr\"ohlich~\cite{prokofev1}
and spin~\cite{prokofev2} polarons. With the same technique the two-body problem was addressed by 
Burovski {\em et al.}~\cite{burovski}
for the exciton problem. Here we work in  direct space (site) representation, the basis consisting
of  Wannier orbitals and phonons at each site. This is in contrast to the exciton problem  where 
the electron-hole interaction is attractive allowing a momentum space calculation
free of the sign problem.
By working in real space we have managed to avoid the sign problem which would appear in the momentum representation 
when the Coulomb repulsion is introduced.
The code can  be easily adapted to include longer range electron-phonon
or electron-electron interactions and to study  models more suitable to the cuprates, 
as for example the extended HH model~\cite{bonca1}.
The disadvantage is that in this basis the momentum dependence of different quantities
is difficult or sometimes impossible to   compute.
We consider a square lattice of $25\times25$ sites with periodic boundary conditions
which is large enough for  negligible finite size errors. 
There are no other truncations of the Hilbert space.


\section{Model Hamiltonian}
\label{sec:modham}

The Hubbard-Holstein Hamiltonian reads 

\begin{eqnarray}
\label{eq:ham}
 H & = &  -t \sum_{\langle ij \rangle, \sigma}( c^{\dag}_{i \sigma} c_{j\sigma} +H.c)
 + U \sum_{i}  n_{i\uparrow} n_{i \downarrow} +\nonumber \\
 & & +\omega_0  \sum_{i} b^{\dag}_{i} b_{i}+  g \sum_{i,\sigma} n_{i \sigma}
 (b^{\dag}_{i}+ b_{i})~~.
\end{eqnarray}

Here $c^{\dag}_{i \sigma} (c_{i \sigma})$ is the creation (annihilation) operator 
of an electron with spin $\sigma$ at site $i$.
$b^{\dag}_{i}$, $ b_{i}$ are phonon creation and respectively annihilation operators at site $i$.
The first term describes the nearest-neighbor hopping of the electrons, and the
second the on-site Coulomb repulsion between two electrons. The lattice degrees of freedom are 
described by a set of independent oscillators at each site, with frequency $\omega_0$.
The electrons couple through the density 
$n_{i \sigma}=c^{\dag}_{i \sigma} c_{i\sigma}$ to the local lattice  displacement
 $x_{i}\propto(b^{\dag}_{i}+ b_{i})$ with a strength $g$.
This Hamiltonian describes a tight-binding model together with an on-site  Coulomb repulsion term
and an on-site electron-phonon interaction term. 
The Holstein and the Hubbard models are limiting cases for $U=0$ and $g=0$ respectively.

In this paper we  address  the electron pairing  as  a function 
of both Coulomb repulsion and  electron-phonon  interaction by studying 
two electrons on a square two-dimensional lattice.

\section{Perturbation theory results}
\label{sec:perturb}

The HH model cannot be solved analytically except for
the two extreme cases of weak and strong electron-phonon interaction, when perturbation theory
can be applied. The most interesting physical situation is in between these regimes.
In order to understand what  happens in the intermediate region, it is necessary to present first
the weak and strong coupling cases.

\subsection{Weak electron-phonon coupling}
\label{sec:weakperturb}

For $g=0$, the ground state will be formed  by two  electrons with zero momentum moving 
freely through the lattice. The total spin is zero because the triplet state could not have
the two electrons in the same $k=0$ state.

When the electron-phonon interaction is switched on, two things happen. The
electrons  get lightly dressed which results in
an  increase of their effective mass, and  the 
electron-phonon interaction introduces a frequency dependent effective attraction between the electrons. 
Up to second order in $g$ the effective attraction is
proportional to  the phonon propagator 
\begin{equation}
\label{eq:wcd}
 V_{eff}^{ph}(\omega)= g^2 \D(q,\omega)=-\frac{2 g^{2}\omega_0}{\omega_0^{2}-\omega^{2}}~~.
\end{equation}

\noindent This is a retarded  interaction and attractive  at small frequency (for $\omega<\omega_0$).
In the antiadiabatic limit ($\omega_0 \longrightarrow\infty$) where the ions are considered 
light and able to follow instantaneously the motion of the electrons, 
the  effective  interaction (Eq.~\ref{eq:wcd}) is instantaneous.

In our model the Coulomb repulsion  competes with the  phonon-induced attraction, 
resulting in a total effective interaction 

\begin{equation}
\label{eq:wcv}
 V_{eff}(\omega)=U-\frac{2 g^{2}\omega_0}{\omega_0^{2}-\omega^{~2}}~~. 
\end{equation}

\noindent In the antiadiabatic limit ($\omega_0 \longrightarrow\infty$)  the  effective  
interaction (Eq.~\ref{eq:wcv})
is instantaneous with
\begin{equation}
\label{eq:wcva}
V_{eff}=U-\frac{2 g^{2}}{\omega_0}
\end{equation} 
\noindent and the situation can be described by a pure  Hubbard model.

The ability of the interaction to bind the electrons into a pair is essential.
It is  well known that in a two-dimensional lattice any attractive instantaneous
interaction will cause  two electrons to  bind in a pair.
We are not aware of any analytical proof that this is true for a retarded interaction (finite phonon
frequency), but a simple numerical
calculation on a variational space which allows states with maximum
one phonon shows that the binding persists  for $\omega_0/t < 1$. 
However, the binding energy is always very small (smaller than $10^{-3}  t$)
and decreases roughly linearly with decreasing $\omega_0$. With increasing $U$, in antiadiabatic limit,
the pairing persists  as long as $U< 2 g^{2}/ \omega_0$.
For finite $\omega_0$
the binding energy decreases  rapidly with increasing $U$. It is  plausible that
the critical $U$ defining the pair formation to be still $2 g^{2}/ \omega_0$, but
the binding energy becomes so small with increasing $U$  that it is very
difficult to resolve  numerically.
However, we do believe that such extremely weak bound pairs have no physical
importance.
Besides, when the electron-phonon coupling is increased 
the situation gets rapidly  complicated and the calculations restricted on the variational
space with a maximum of one phonon are inappropriate.
The renormalization of
the electron-phonon interaction vertex becomes important. Migdal's theorem~\cite{migdal} applied 
in classical superconductivity theory  is not valid here because of  the absence of the 
Fermi sea\footnote{Migdal's theorem shows that the vertex corrections are of order $\omega_0/\E_F$ due to Pauli exclusion principle which blocks the electron-phonon
scattering inside the Fermi sea.}.


\subsection{Strong electron-phonon coupling}
\label{sec:strongperturb}

In the strong coupling limit, the HH model may be addressed with a canonical 
transformation and  by treating the hopping part of the Hamiltonian as a 
perturbation.  
The last three terms in Eq.~(\ref{eq:ham}) are diagonal in 
the rotated basis obtained by applying the unitary operator~\cite{LF} $e^{\textstyle S}$ where
\begin{equation}
\label{eq:lft}
S=-\frac{g}{\omega_0} \sum_{i,\sigma} n_{i \sigma}(b^{\dag}_{i}- b_{i})~~.
\end{equation}

\noindent Using formula 
\begin{equation}
\label{eq:ut}
\tilde{A}=e^{S} A  e^{-S}=A+[S,A]+\frac{1}{2}[S,[S,A]]+ ..
\end{equation}

\noindent the transformed operators become
\begin{eqnarray}
\label{eq:troppol}
\tilde{b}_{i}=b_{i}+\sum_{\sigma} \frac{g}{\omega_0} n_{i \sigma}  \\
\tilde{c}_{i \sigma}=c_{i \sigma}\ e^{ \textstyle \frac{g}{\omega_0}(b^{\dag}_{i}- b_{i})}~~.
\end{eqnarray}

\noindent The Hamiltonian written in the new basis is
\begin{equation}
\label{eq:nham}
 H=H_{t} + H_{0} 
\end{equation}
\noindent with
\begin{equation}
\label{eq:nham0}
H_{0}= \omega_0  \sum_{i} \t{b}^{\dag}_{i} \t{b}_{i}- \frac{g^{2}}{\omega_0} \sum_{i,
\sigma} \t{n}_{i,\sigma} +(U-\frac{2 g^{2}}{\omega_0})\sum_{i}\t{n}_{i\uparrow}
\t{n}_{i \downarrow} 
\end{equation}
\begin{equation}
\label{eq:nhamt}
H_{t}= -t \sum_{\langle ij \rangle, \sigma}( \t{c}^{\dag}_{i \sigma} \t{c}_{j\sigma} 
X^{\dag}_{i} X_{j} +H.c) 
\end{equation}
\noindent and
\begin{equation}
\label{eq:xxpol} 
X_{i}=e^{ \textstyle -\frac{g}{\omega_0}(\t{b}^{\dag}_{i}- \t{b}_{i})}~~.
\end{equation}

The physical meaning of this canonical transformation is a shift of the
ions  equilibrium position at the sites where the electrons are present

\begin{equation}
\label{eq:ionpos}
\langle \tilde{x_i} \rangle = \langle \tilde{b}^{\dag}_i +\tilde{b}_i \rangle = 
\langle b^{\dag}_i +b_i + \sum_{\sigma} \frac{2g}{\omega_0} n_{i\sigma}\rangle = \langle x_i\rangle +
\frac{2g}{\omega_0} \langle n_i\rangle ~~.
\end{equation}

\noindent As can be seen from the second term of Eq.~\ref{eq:nham0}, the lattice deformation energy gained 
due to the electron  presence is
\begin{equation}
\label{eq:eppol} 
E_{p}= g^{2}/\omega_0 ~~.
\end{equation}

\noindent  The dimensionless  electron-phonon coupling constant may be defined as the ratio 
between this energy and the bare electron kinetic energy which is proportional to the hopping $t$ 
and with the lattice dimensionality $z$.
We define it as  
\footnote{Other authors define the dimensionless coupling constant as the ratio between the deformation energy 
and the phonon
frequency~\cite{kittel}, which will be a measure of the number of phonons in the polaron cloud.\\
If in BCS theory of superconductivity the approximation $N(0) \approx 1/2W$ is made, where $N(0)$ is the density of
states at the Fermi level and $W$ is the electron bandwidth,
the relation of our $\alpha$ to the BCS electron-phonon coupling,  $\lambda=2 g^{2}/\omega_0 N(0)$, will 
be $\alpha=2\lambda$.} 

\begin{equation}
\label{eq:alpha1} 
\alpha=  \frac{g^{2}}{\omega_0 zt} =\frac{g^{2}}{2\omega_0 t} ~~.
\end{equation}

\noindent Since the electron hopping is accompanied by  a change in  ions equilibrium
position (see the term $X^{\dag}_{i} X_{j}$ in Eq.~\ref{eq:nhamt}),  it is exponentially reduced 
for large $g$.
\begin{equation}
\label{eq:teff}
t_{eff}= t~\langle i| \t{c}^{\dag}_{i } \t{c}_{j} 
X^{\dag}_{i} X_{j} | j \rangle = t ~e^{\textstyle{-\frac{g^2}{\omega^2_0}}} =
t ~e^{\textstyle{-\frac{\alpha z t}{\omega_0}}}
\end{equation}

\noindent The effective on-site interaction between electrons is 
\begin{equation}
\label{eq:u1}
U_{eff}=U-2\frac{g^{2}}{\omega_0} \equiv U- 2 E_p
\end{equation} 
\noindent (the same as in Eq.~(\ref{eq:wcva})), 
and in the antiadiabatic limit (when $\omega_0,g\longrightarrow\infty$, $g/\omega_0\longrightarrow 0$ and
$2g^{2}/\omega_0$ is finite) $X_i=1$ and the model can be mapped again in a pure Hubbard one.

For negative  $U_{eff}$ it is evident that the electrons form a bound state.
However, based on  second order perturbation theory in the hopping  $H_t$, it can be shown 
that  even for positive $U_{eff}$ a stable bipolaron state can exist.
Let's consider the case of large $U$ which results in $U_{eff}>0$.
The ground state of $H_0$ is formed by the degenerate states
\begin{equation}
\label{eq:gs}
|a_{i}\rangle= c^{\dag}_{i \uparrow}c^{\dag}_{i+a \downarrow}|0 \rangle,~~a\neq 0~~.
\end{equation}
The  meaning of this notation is that the electron with spin $\downarrow$ is at a  distance
 ``$a$'' from the one  with spin $\uparrow$ residing at site ``$i$''.
``$a$'' can take all the possible
 values except $0$.
In first order perturbation theory, the calculation of the matrix $\langle a_{i} \mid H_t \mid b_{j} \rangle$ 
reduces to the calculation of Eq.~\ref{eq:teff} and indicates an
 exponentially reduced nearest-neighbor hopping.
Second order perturbation theory stabilizes the bipolaron states. It is equivalent to 
diagonalizing the operator
\begin{equation}
\label{eq:T}
T=H_{t} \frac{1}{E_{0}-H_{0}} H_{t}
\end{equation}
\noindent on the subspace spanned by all the degenerate states of $H_{0}$.
\begin{figure}[t]
\centerline{
\includegraphics*[width=3.3in]{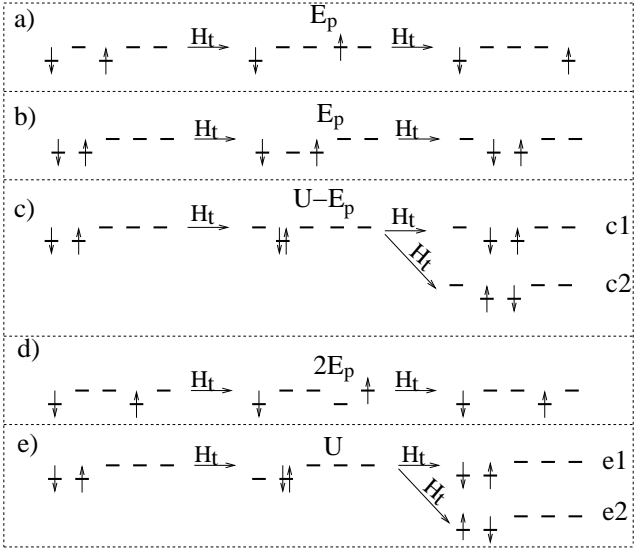}}
\caption{Schematic representation of second order processes in $H_t$ which 
determine the elements of matrix $T$. The
small horizontal lines represent the lattice sites. If an electron
is on a particular site then the equilibrium position of the ion at that site will
be changed, and we indicate this by a downwards shift. 
The excitation energy of the intermediate state is shown  above every process.
If the initial and the final lattice configurations are different, the  corresponding matrix
 element will be exponentially reduced. In the large $g$ limit 
the corresponding matrix elements are:
a) $-\frac{\textstyle t^2}{\textstyle E_p} e^{\textstyle -g^2/\omega_0^2}$. 
b) $-\frac{\textstyle t^2}{\textstyle E_p} e^{\textstyle -g^2/\omega_0^2}$.
c) $-\frac{\textstyle t^2}{\textstyle U-E_p} e^{\textstyle -g^2/\omega_0^2}$.
d) $-\frac{\textstyle t^2}{\textstyle 2 E_p}$.
e) $-\frac{\textstyle t^2}{\textstyle U }$.} 
\label{fig:sop}
\end{figure}
The processes which can take place are shown schematically in Fig.~\ref{fig:sop}.
We can classify them in two classes. The first class includes processes as in  Fig.~\ref{fig:sop}-a,
-b and -c, where the final lattice  configuration is different from
the initial one.  This results in an exponential reduction of the matrix elements,
so that we can neglect them in  first approximation. 
The second class includes
 processes like the ones in Fig.~\ref{fig:sop} -d and -e where the initial and the final
lattice configuration is unchanged. They are not exponentially reduced.
In Fig.~\ref{fig:sop}-d an electron hops on a neighboring site without carrying the lattice deformation 
around it and afterwards  comes back. The energy of the intermediate state is $2E_{p}$,
because it contains a site with deformation and without electron, and a site with an electron and without 
deformation. The gain in energy is  $-t^{2}/2E_{p}$. This process only contributes to the diagonal 
elements of $T$.
In Fig.~\ref{fig:sop}-e one electron hops without carrying the lattice deformation 
on the neighboring site which is occupied by the other electron.
The intermediate state contains  a doubly occupied site which has a deformation corresponding
to only one electron and a site with deformation and without electron, therefore the
energy of this state is  $U_{eff}+2E_{p}=U$. 
The final state can be identical with the initial state (Fig.~\ref{fig:sop}-e1), and the process contributes to the 
diagonal elements
$T_{\delta,\delta}$,
or  the final state can have the electrons  interchanged (e.g.: from the initial
$\mid \uparrow,\downarrow>$ to the final $\mid \downarrow,\uparrow>$ - Fig.~\ref{fig:sop}-e2), 
and this process contributes 
to the non-diagonal elements $T_{\delta,-\delta}$. $\delta$ means nearest-neighbor here. 
The energy gain corresponding to each of the two  processes shown in  Fig.~\ref{fig:sop}-e is  $-t^{2}/U$. 
Neglecting the exponentially reduced terms in the calculation of $T$, what remains is the 
diagonal terms and the off-diagonal ones which connect the states $\mid \delta\rangle$
and $\mid -\delta\rangle$. 
\begin{eqnarray}
\label{eq:tvalue}
\begin{array}{lclr}
T_{a,a}&= & 8 \times (-\frac{\textstyle{t^2}}{\textstyle{2E_p}}) & a\neq 0,1  \\
T_{\delta,\delta}&= &6\times (-\frac{\textstyle{t^2}}{\textstyle{2E_p}})+ 2\times (-\frac{\textstyle{t^2}}{\textstyle{U}})&\\
T_{\delta, -\delta}&= & 2\times (-\frac{\textstyle{t^2}}{\textstyle{U}}) & \delta=1
\end{array}
\end{eqnarray}
Solving the secular equation, we find  the condition for the  bipolaron existence  to be
\begin{equation}
\label{eq:condb}
 U < 4 E_{p}
\end{equation}
\noindent and the bipolaron binding energy 
\begin{equation}
\label{eq:Deltab}
\Delta_b =-\frac{t^2}{E_p}+ \frac{4t^2}{U} ~~.
\end{equation}
Notice that even though the bipolaron exists up to a large value of $U$ it is a weakly bound state
when $U_{eff}>0$.

The physical interpretation of these results is simple.
The energy given by Eq.~\ref{eq:Deltab} corresponds to a double degenerate singlet state formed by  
two electrons on 
nearest-neighbor sites. One state is a singlet along the $X$ direction and the other along the
$Y$ direction.  In distinction
to Hubbard model where the exchange energy can never win over the kinetic energy and
therefore cannot bind two electrons, here the interaction with phonons results in a strong
reduction of the electron bandwidth but not of the exchange energy because it implies
virtual transitions of  electrons on double occupied sites 
{\em without carrying the lattice deformation} with them. Therefore  now the exchange energy can
easily win and produce singlet bound states. However there is another effect which
introduces an effective repulsion between two nearest-neighbor electrons and 
wins over the exchange energy when $U \geq 4 E_p$. A virtual transition of an
electron to an empty nearest-neighbor site without carrying the lattice deformation
will lower its energy by  $-t^2/2 E_p$ (Fig.~\ref{fig:sop}-d). But if the nearest-neighbor
site is occupied by the other electron this process is not possible 
resulting in an effective repulsion  of $t^2/E_p$ between two nearest-neighbor electrons. 
Therefore Eq.~\ref{eq:Deltab} reflects the competition between this effective repulsion and the exchange attraction
equal to $-4t^2/U$.

When the processes shown in  Fig.~\ref{fig:sop} -$a$, -$b$ and -$c$ are taken into account 
 the degeneracy of the two singlets is lifted, and two states are formed.
This results in a ground state with $s$-wave ($A_{1g})$ symmetry and another state with $d$-wave ($B_{1g}$) 
symmetry.  It should also be mentioned that 
if a positive next-nearest-neighbor hopping $t'$ is introduced in the model, the $d$-wave symmetry state 
will be stabilized and it  becomes the ground state when $t'$ is large enough.

Let's summarize the strong coupling regime physics, neglecting at the beginning 
the exponentially reduced terms.
When $U$ is small the ground state energy is  $U-4 E_p$ and consists
of two electrons  located on the same site. 
The first excited state contains one more phonon and has an energy $U-4 E_p+\omega_0$ 
(this is a $N$ degenerate state because the phonon can be at any site).
When $U$ is increased and $U-4 E_p+\omega_0$ becomes larger than $-2 E_p$ (which is the zero order energy of
two electrons staying on different sites), the first excited state
is a double degenerate nearest-neighbor singlet.
When the hopping is switched on the low-energy physics can be described by the Hamiltonian
\begin{eqnarray}
\label{eq:hammap}
 H & = & -t_{eff} \sum_{\langle i,j \rangle, \sigma}( c^{\dag}_{i \sigma} c_{j\sigma} +H.c)
  + J \sum_{\langle ij\rangle} 
({\bf S}_{i} {\bf S}_{j}-\frac{n_in_j}{4})  \nonumber \\ 
 & + & V \sum_{\langle i,j \rangle}  n_i n_j +
U_{eff} \sum_{i}  n_{i\uparrow} n_{i \downarrow} + H'
\end{eqnarray}
\noindent with the  hopping $t_{eff}= t~ e^{\textstyle -g^2/\omega_0^2}$, 
the  exchange $J= \frac{\textstyle 4 t^2}{\textstyle U}$,
the nearest-neighbor repulsion $V= \frac{\textstyle t^2}{\textstyle E_p}$ and the
on-site interaction $U_{eff}=U-2 E_p$.
$H'$ describes the processes shown in  Fig.~\ref{fig:sop} -$a$, -$b$ and -$c$. Their magnitude
is either  $\frac{\textstyle t_{eff}}{\textstyle E_p}$ or 
$\frac{\textstyle t_{eff}}{\textstyle U-E_p}$ (see the caption of Fig.~\ref{fig:sop}),
which is  much smaller than $t_{eff}$.
In the  literature~\cite{bonca,aubry} the bipolaron with the electrons located on the same site is called $S0$,
and the one with the electrons  located on nearest-neighbor sites $S1$. 
The Hamiltonian~(\ref{eq:hammap}) describes the transition from the $S0$ to the  $S1$ bipolaron in strong 
coupling regime.
For small (i.e. negative) $U_{eff}$ the order of the lowest energy states is: $s$-wave $S0$, 
$s$-wave $S1$, $d$-wave $S1$.
When $U_{eff}$ increases the $S0$ state starts mixing with $s$-wave $S1$ state.
The mixing between the  $S0$ and  the $s$-wave $S1$ states is of  order of $t_{eff}$,
and the splitting 
between the $s$-wave and the $d$-wave $S1$ states is given by $H'$, thus being much smaller.
The order of low-energy states  becomes: linear combination of $s$-wave $S0$ and $S1$,
$d$-wave $S1$, linear combination of $s$-wave $S0$ and $S1$. 
For larger $U$ only two bound states exists: s-wave $S1$ and d-wave $S1$. 
In conclusion, the ground state evolves analytically (crossover) from $S0$ bipolaron to $s$-wave $S1$ bipolaron with
increasing $U$. The situation is different for the first excited state.
Here at a critical value of $U$, a nonanalytical transition takes place, and the 
first excited state changes from   $s$-wave symmetry to $d$-wave symmetry.

\section{Algorithm}
\label{sec:algorithm}

\subsection{General approach}
\label{subsec:gi}

Our algorithm calculates different imaginary time Green's functions and
relies upon the ability to project out 
the ground state properties by extrapolating to long complex times. 
Let's consider the equation
\begin{equation}
\label{eq:polal1}
 \langle \psi | e^{\textstyle{-\tau H}} | \psi \rangle = \sum_{\nu} |\langle \psi |\nu \rangle|^{2} e^{\textstyle{-\tau \e_{\nu}}}
\end{equation}
\noindent where $|\psi \rangle $ is a whatever state and $ \{ |\nu  \rangle \}$ form the complete set of the eigenstates
with energies $\e_\nu$.
We see that at large $\tau$   Eq.~\ref{eq:polal1} converges to 
\begin{equation}
\label{eq:polal2}
| \langle\nu_0 |  \psi\rangle|^{2} e^{\textstyle{-\tau \e_{\nu_0}}}
\end{equation}
\noindent where $ |\nu_0 \rangle$ is the ground state of the system.
Suppose the ground state is separated from the first excited state by a gap
$\Delta$. We can obtain the ground state energy and the overlap of the ground state with $|\psi \rangle $
with an accuracy better than 1\% (for example)  calculating  Eq.~\ref{eq:polal1} at a time 
$ \tau \approx 5/ \Delta $.

Because the total momentum  $K$ is a quantity which is conserved in our problem
we can obtain the lowest energy in the $K$ channel by calculating
\begin{widetext}
\begin{eqnarray}
\label{eq:bipal3}
 P^{n}(K,\tau) =  
 \sum_{k,q_{1},...q_{n}}\langle (K-k-q_{1}-...-q_{n})_{\downarrow},k_{\uparrow};q_{1},...,q_{n}|
 e^{-\tau H}|(K-k-q_{1}-...-q_{n})_{\downarrow},k_{\uparrow};q_{1},...,q_{n} \rangle 
\nonumber\\
 \longrightarrow \sum_{k,q_{1},..q_{n}}
 | \langle (K-k-q_{1}-...-q_{n})_{\downarrow},k_{\uparrow};q_{1},...,q_{n}|\nu_{0K} \rangle |^{2}
 e^{-\tau E(K)}\nonumber\\
\end{eqnarray}
\end{widetext}
\noindent at large $\tau$. Here $|k_{1 \downarrow},k_{2 \uparrow};q_{1},q_{2},...,q_{n} \rangle $
is a state with two electrons, one with momentum $k_{1}$ and spin down and the other
with momentum $k_{2}$ and spin up, and with $n$ phonons with momentum $q_{1}$, $q_{2}$,... and $q_{n}$
respectively. $| \nu_{0K} \rangle $ is the ground state (the state with the lowest energy) in the  $K$ channel.
The calculation of  $P^{n}(K,\tau)$  yields both the bipolaron energy and the  $n$-phonon
configuration probability in the bipolaron state.

For reasons related with the sign problem 
(to be discussed later), we  calculate  $P^{n}(K,\tau)$ in
real-space representation

\begin{equation}
\label{eq:al4}
P^{n}(K,\tau)=\frac{1}{N}\sum_{i,x,l_{1},l_{2},...l_{n}}e^{i K x} \langle i | e^{-\tau H} T_{x}|i \rangle 
\end{equation}
\noindent where 
\begin{equation}
\label{eq:al5}
|i \rangle \equiv | i_{\downarrow},(i+a)_{\uparrow};i+l_{1},i+l_{2},...,i+l_{n} \rangle
\end{equation}
\noindent is a state 
with a spin down electron at site $i$, a  spin up electron at site $i+a$ and phonons at sites $i+l_{1}$, $i+l_{2}$,
... and $i+l_{n}$,
and 
\begin{eqnarray}
\label{eq:al6}
& T_{x}|i \rangle  = & \nonumber\\ 
&| i+x_{\downarrow},(i+x+a)_{\uparrow};i+x+l_{1},i+x+l_{2},...,i+x+l_{n}\rangle & \nonumber\\ 
\end{eqnarray}
\noindent is the state $|i \rangle$ translated with the vector $x$.

Another quantity which is conserved is the total spin. 
Therefore for the singlet we calculate
\begin{equation}
\label{eq:bipal7}
P^{n}_{s}(K,\tau)=\frac{1}{N}\sum_{i,x,l_{1},l_{2},...l_{n}}e^{i K x} \langle i_{s} | e^{-\tau H} T_{x}|i_{s} \rangle 
\end{equation} 
\noindent with
\begin{equation}
\label{eq:al8}
|i_s \rangle \equiv | (i,i+a)_s;i+l_{1},i+l_{2},...,i+l_{n} \rangle
\end{equation}
\noindent where $(i,i+a)_s$ is the singlet state with electrons at sites $i$ and $i+a$.
Similar equations can be written for the triplet channel.

In order to calculate the above  quantities, we developed a  DQMC code 
which  stochastically generates terms of the form
\begin{equation}
\label{eq:bipqmc1}
G_{ij}(\tau)= \langle i | e^{-\tau H} | j(i,x) \rangle 
\end{equation}
\noindent where  $| i \rangle $ is a general state as in Eq.~\ref{eq:al5} with
two electrons at  arbitrary
distance from each other and  with an arbitrary number of phonons. The state 
$| j (i,x)\rangle $ can be obtain either by applying a translation operation with an arbitrary vector $x$
on $| i \rangle $ (i.e.  \mbox{$ | j \rangle = T_{x}  | i_{\downarrow}, (i+a)_{\uparrow}; phonons \rangle$})
or  by applying a permutation (interchanging the electrons position) and a translation  on $| i \rangle $
(i.e. \mbox{$| j \rangle = T_{x}  | (i+a)_{\downarrow}, (i)_{\uparrow}; phonons \rangle$}).

The value of an observable $A$ in a particular
$K$ and $S$ channel is
\begin{eqnarray}
\label{eq:bipqmc2}
 A(K)  = \frac{1}{M}\sum_{m} e^{iKx(m)} g^{S}(m) a(m)=\nonumber\\
=\frac{\sum_{m} e^{iKx(m)} g^{S}(m)w(m)a(m)}{\sum_{m}w(m)}
\end{eqnarray}

\noindent In Eq.~\ref{eq:bipqmc2}  we sum over all $m$  generated configurations with the weight $w(m)$. 
$M$ is the total number of measurements,
$x(m)$ is the translation vector which correspond to the configuration $m$, $a(m)$ is the estimator of
$A$  and $g^{S}(m)$ is
the factor which separates the triplet from the singlet.
For singlet $g^{S}(m)$ is $1$ when electrons are on the same site and
$1/2$ otherwise. For triplet  $g^{S}(m)$ is zero when the electrons are on the same site,
and otherwise, $1/2$ when  $| j \rangle $ is a translation of  $| i \rangle $ and $-1/2$
when  $| j \rangle $ is a translation of a state obtained from $| i \rangle $ by interchanging
the electrons position.

\subsection{Implementation}
\label{bip:dqmcimp}

Let's start from the Hamiltonian~(\ref{eq:ham}), and consider 

\begin{eqnarray}
\label{eq:qmcham}
H_{0}=\omega_0  \sum_{i}  b^{\dag}_{i} b_{i} + U \sum_{i}  n_{i\uparrow} n_{i \downarrow}
\end{eqnarray}
as the noninteracting part of the Hamiltonian. $H_{0}$ is diagonal in the real space
representation.
The evolution operator can be written as
\begin{eqnarray}
\label{eq:qmcev}
 e^{-\tau H}=e^{-\tau H_{0}} S(\tau) 
\end{eqnarray}
with
\begin{eqnarray}
\label{eq:qmcs}
 S(\tau)= \sum_{n=0}^{\infty} \frac{(-1)^{n}}{n!} \int_{0}^{\tau}...\int_{0}^{\tau}
d\tau_{1}...d\tau_{n} T[H_{1}(\tau_{1})...H_{1}(\tau_{n})] \nonumber \\
\end{eqnarray}
where $H_{1}=H-H_{0}$  in the interaction picture is
\begin{eqnarray}
\label{eq:qmcham1}
&  H_{1}(\tau) & = e^{\tau H_{0}} H_{1} e^{-\tau H_{0}}~~.\nonumber\\	
\end{eqnarray}
 
Eq.~\ref{eq:bipqmc1} becomes:

\begin{widetext}
\begin{eqnarray}
\label{eq:qmcij}
G_{ij}(\tau)= \langle i | e^{-\tau H} | j \rangle 
= e^{-\E_{i} \tau}\sum_{n=0}^{\infty} \frac{(-1)^{n}}{n!} \int_{0}^{\tau}\int_{0}^{\tau}...\int_{0}^{\tau}
d\tau_{1}...d\tau_{n}T[ \langle i|H_{1}(\tau_{1})H_{1}(\tau_{2})...H_{1}(\tau_{n})| j \rangle] 
\end{eqnarray}
\end{widetext}

\begin{figure}[t]
\includegraphics*[width=3.3in]{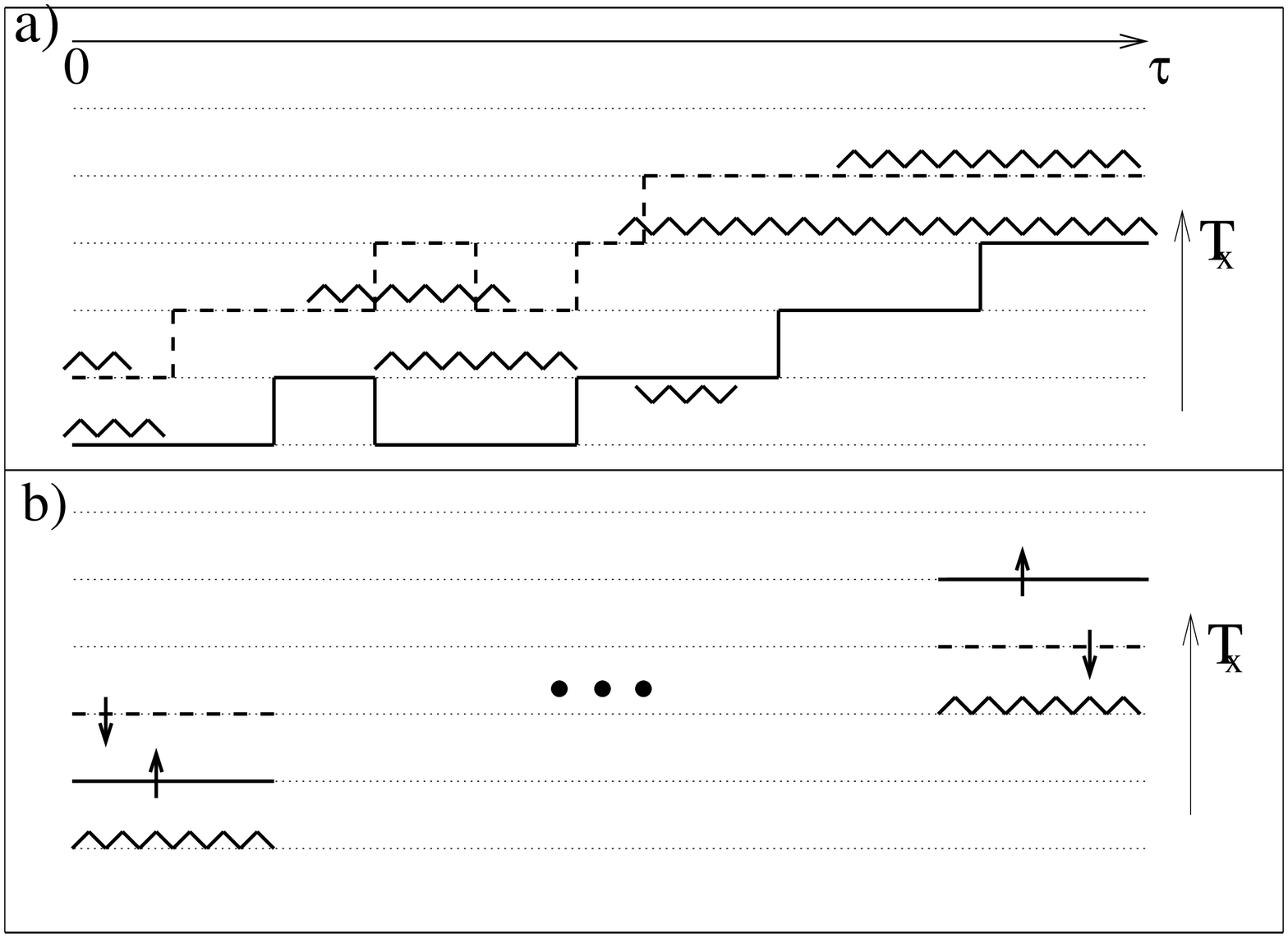}
\caption{a) A typical diagram which represents a term in Eq.~(\ref{eq:qmcij}).
The solid line (dashed line)  represents a spin up (down) electron. The wavy
line is a phonon propagator. The end at time $\tau$ is a translation with the 
vector $x$ of the end at time 0. 
b) The final state  is a translation of the state obtained from initial state with 
the electrons position interchanged.
In the code we consider only configurations where the final state is a translation
of the initial state as in a) or is a translation of the state obtained from initial state with 
the electrons position interchanged as in b).}
\label{fig:mcdiag}
\end{figure}

\begin{figure}[t]
\centerline{
\includegraphics*[width=3.3in]{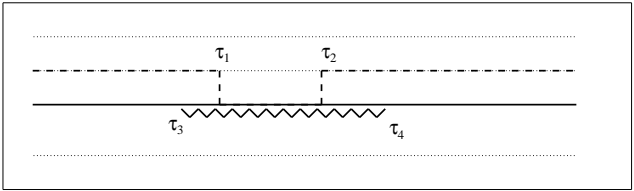}}
\caption{ The weight of this diagram is $(t d \tau )^2 ~ e^{\textstyle -U(\tau_{2}-\tau_{1})}
~ (g d \tau) ^2 ~ e^{\textstyle -\omega(\tau_{4}-\tau_{3})} $
The weight of a diagram is determined by the following rules:
(i) every electron hopping corresponds to a term $t~d\tau$,
(ii) every electron-phonon vertex corresponds to a term $g~ d\tau$,
(iii) every phonon propagator of length $\Delta \tau$ corresponds to a term $e^{\textstyle{-\omega_0 \Delta \tau}}$,
and (iv) every interval where the electrons are  on the same site during a time $\Delta \tau$ 
corresponds to a term $e^{\textstyle{-U \Delta \tau}}$.}
\label{fig:mcrules}
\end{figure}

The calculation of $G_{ij}(\tau)$ is reduced to
a series of integrals, with an ever increasing number of integration variables.
It is easy to show that every term in Eq.~(\ref{eq:qmcij}) can be represented by a diagram
and a set of simple rules can be derived to determine the diagrams weight.
Typical examples of  such diagrams are presented in  Fig.~\ref{fig:mcdiag}. 
Aside from a translation, the electronic configuration at the diagram ends must
be either identical (Fig.~\ref{fig:mcdiag}-a) or with the electrons position interchanged
(Fig.~\ref{fig:mcdiag}-b). The rules for determining the diagram weight are given in the caption 
of Fig.~\ref{fig:mcrules}.

We generate all the possible diagrams with an arbitrary number of phonons,
and with the difference between $0$ and a $\tau_{max}$ chosen large enough to project 
the ground state. Our code is a continuous time code (i.e. it does not require artificial discretization of
the imaginary time axis), and the diagrams are generated is a manner similar  
to that described in~\cite{prokofev,prokofev1,burovski}. 

Estimators for energy, effective mass, phonon distribution and correlation function of
the electrons position can be easily found. The measurements are taken only
at large time where the ground state is projected out.
The bipolaron energy estimator is~\cite{prokofev}
\begin{equation}
\label{eq:este}
E(m)=\frac{1}{\tau_m}(\omega_0\sum_{i_{ph}} \tau_{i_{ph}}+U\sum_{j_{U}}\tau_{j_{U}}-N_{vertex}-N_{hop})
\end{equation}
where $\tau_m$ is the time (length) of the $m$ diagram, $i_{ph}$ counts the phonons propagators 
with $\tau_{i_{ph}}$ length,
$j_{U}$ counts the intervals with  double occupied sites  with $\tau_{j_{U}}$ length, $N_{vertex}$ is the
number of electron-phonon vertices and $N_{hop}$ is the number of electron hopping jumps.
The estimator for the inverse of the bipolaron effective mass is
\begin{equation}
\label{eq:estm}
\frac{2 m_e}{m^{*}}(m)=\frac{x(m)^{2}}{\tau_m}
\end{equation}
where $x(m)$ is the translation vector between the time $0$ end and the time $\tau_m$ end. $m_e$
is the free electron effective mass.
The probability to have $n$ phonons is calculated with the estimator
\begin{equation}
\label{eq:estph}
z^n(m)=\delta_{n_m,n}
\end{equation}
where $n_m$ is the number of phonons at the ends of the diagram.
The electrons relative positions correlation function defined as
\begin{equation}
\label{eq:corr}
C(r)= \frac{1}{N}\sum_{i}\langle n_{i} n_{i+r} \rangle
\end{equation}
has the estimator
\begin{equation}
\label{eq:estcorr}
C(r;m)=\delta_{r,R_m}
\end{equation}
where $R_m$ is the relative distance between electrons at the ends of the measured diagram.

\begin{figure}[t]
\centerline{
\includegraphics*[width=3.3in]{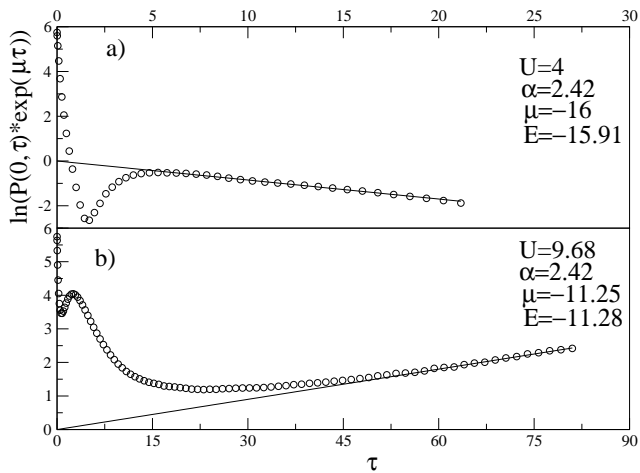}}
\caption{ $ln ( P_{s}(0,\tau)*e^{\textstyle \mu \tau}) $ versus $\tau$. a) For the given parameters a large bipolaron
binding energy results. b) For the given parameters  a small bipolaron
binding energy results. The linear asymptotic behavior starts at smaller time in case a).
Notice that the time scale is different for the 
two cases presented.}
\label{fig:green}
\end{figure}

As an illustration of our method, in Fig.~\ref{fig:green} we show $ln(P_s(0,\tau)*e^{\textstyle \mu \tau})$ 
versus $\tau$ where
\begin{equation}
\label{eq:z0}
P_{s}(0,\tau)*e^{\mu \tau}=\sum_{n}P^{n}_s(0,\tau)*e^{\mu \tau} = \sum_{\nu} e^{-(E_{\nu}-\mu)\tau}
\end{equation}
\noindent with $P^{n}_{s}$  defined in Eq.~(\ref{eq:bipal7}) for the $K=0$  channel. 
$\mu$ is an arbitrary parameter  which is chosen close to the bipolaron energy to avoid
the exponentially small weight of the large time diagrams.
It can be seen that at long imaginary time  $ln(P_{s}(0,\tau))$ becomes linear in $\tau$, the
slope being proportional to the ground state energy. 
An important remark should
be made about the strong  drop seen in $P_{s}(0,\tau)$  at short time. This is due to the fact that
we generate only connected diagrams (diagrams where the phonon propagators
are always glued to the electron propagator). The disconnected diagrams have an exponentially small contribution
at large time, therefore they can be safely neglected, but at small time their omission will result in
a strong potential drop which will not allow an  efficient sampling  for both  
long  and short time diagrams. We have eliminated this problem using a fictitious potential 
renormalization~\cite{prokofev}.

\subsection{Discussions}
\label{bip:dqmcdis}

In order to avoid the sign problem,
our code calculates the Green's functions in the real-space representation (Wannier basis),
even though  this representation has some  disadvantages.
A similar expression to Eq.~\ref{eq:qmcij} can be written in the momentum space (Bloch basis), 
and similar rules for determining the  diagrams weight can be found. This approach was considered
in~\cite{burovski} for the exciton model calculation. 
The problem with our model is that, unlike  in the exciton case where the 
conduction-electron valence-hole interaction is 
attractive, we have  a repulsive interaction. This will make all the diagrams with
an odd number of electron-electron interaction vertices  negative, which implies a very severe sign problem.
In the real-space representation the sign problem is avoided, all the diagrams being positive definite.
However, this representation introduces other problems, for example it makes the study of the bipolaron at
large momentum difficult and inaccurate. In a more general sense, problems appear in all
the irreducible channels aside from the one which contains the system ground state (i.e. $K=0$, singlet).
The difficulty  is two-fold. First there is the sign problem. Aside  from the  singlet and $K=0$ channel
where all the terms in Eq.~\ref{eq:bipqmc2} are positive definite, at $K \neq 0$ or/and in the triplet 
channel the factors $e^{\textstyle iKx(m)}$ and $g^{S}(m)$ can take negative values.
Second, in the real-space representation we generate all the possible configurations
with all the possible symmetries. To project out the lowest energy state of the channel ``$\gamma$'' 
we have to calculate  $P_{\gamma}(\tau)$ up to a time proportional to $1/\Delta_{\gamma}$, where
$\Delta_{\gamma}$ is the  $\gamma$ channel gap. Therefore we have to simulate  larger imaginary
times for a channel characterized by a smaller gap. But because
all the symmetry channels configurations are generated in the same run,
the statistics for the channel $\gamma$ is proportional to $e^{\textstyle -(E_{\gamma}-E_{0}) \tau}$ 
(here $E_{0}$ is the ground state energy).
Thus if the imaginary time 
is increased, the statistics for channels other than the one which contains the ground
state will be exponentially reduced.

Another problem, specific to all ground state projective algorithms,  occurs if there are more than one bound state in
the same symmetry channel, quasi-degenerate in energy. In this case, at large imaginary 
time we project out all these states.
The results obtained in this case are going to reflect the average of the properties of all
the projected states. We  encounter this problem at large electron-phonon coupling,
where the difference of the $s$-wave and $d$-wave bipolaron energies is exponentially small,
as the strong coupling theory predicts.
However in the intermediate coupling regime we managed to separate  these states and we always 
found a $s$-wave ground state.

Before  we discuss the necessary modifications for adapting our algorithm to other bipolaron models, we 
mention that, even for the  present HH bipolaron, the  code can be improved.
The momentum $K$ and the spin $S$ are not all the
quantum labels which can be used to distinguish between the different symmetry channels. We also have the point 
group symmetries
which break the Hilbert space in different irreducible representations. We have already discussed about
$s$-wave ($A_{1g}$) and $d$-wave ($B_{1g}$) bipolaron states. In principle we can look for all the
symmetries given by the representations of the $D_{4h}$ point group. These symmetries can be separated in a
similar way to what we did when separating the  singlet and the triplet channels. 
We have to  generate diagrams where the  time $\tau$  is obtained after a translation and a point group 
operation applied to the time $0$. In other words the electronic configurations of the diagrams ends 
should be connected by a 
space group operation\footnote{In fact the separation of singlet and triplet obeys the same idea,
the permutation group operations being used.}. For every operation there is a certain factor $t^{D}(m)$ which separates
the different representations, and the value of an observable  is calculated analogues to Eq.~\ref{eq:bipqmc2}
\begin{eqnarray}
\label{eq:bibqmc3}
A(K)  = \frac{1}{M}\sum_{m} e^{iKx(m)} g^{S}(m) t^{D}(m) a(m)=\nonumber\\
=\frac{\sum_{m} e^{iKx(m)} g^{S}(m) t^{D}(m) w(m)a(m)}{\sum_{m}w(m)}
\end{eqnarray}
\noindent For example, at $K=0$,  $t^{D}$ is always $1$ for $A_{1g}$ representation. In general  $t^{D}$
should be proportional to the characteristic of the representation. 
The sign problem can intervene for other representations than $A_{1g}$.
The above approach applied to our present model will improve the accuracy
of the results which describe the properties of the $s$-wave $S1$ bipolaron.
However the study of the $d$-wave symmetry bipolaron will still be difficult and not very accurate,
because of the smallness of the binding energy of this state. 
We believe that no new physics will appear to justify the  large coding effort necessary for implementing the 
above approach.
Nevertheless the separation of the different point group representations could be essential for other 
models which include longer
electron-phonon interaction and where strongly bound bipolarons with electrons residing 
on different sites exist~\cite{alexandrov}.

The code can be easily modified to include longer range electron-electron and electron-phonon interaction.
For an electron-phonon interaction term of the form
\begin{equation}
\label{eq:qmc4}
\sum_{i,j}g_{ij}n_{i}(b^{\dagger}_{j}+b_{j})
\end{equation}
and for an electron-electron interaction  of the form
\begin{equation}
\label{eq:qmc5}
\sum_{i,j}V_{ij}n_{i} n_{j}
\end{equation}
there will be no sign problem. The diagrams are  similar to the ones shown in Fig.~\ref{fig:mcdiag}
aside  from the possibility that now a phonon propagator at site ``$j$'' 
can be created or destroyed by an electron at site ``$i$''.
This  process will have a probability equal to $g_{ij} ~d \tau$.
If the two electrons are on sites 
``$i$'' and ``$j$'' for an interval of  time $\tau$ a term 
$e^{\textstyle{-V_{ij} \tau}}$ in the diagram weight will correspond.

\section{Results}
\label{bip:results}

In all the subsequent calculations the phonon energy and the electron hopping term are chosen to be 
one ($\omega_0=1$, $t=1$). In real materials $\omega_0$ is smaller than $t$ (by almost one order
of magnitude), but  calculations with such small $\omega_0$ are very  time
consuming and  we find  that they do not  provide any new qualitative results. 
Our calculation is done on a $25 \times 25$ square lattice with periodic boundary conditions.

\subsection {Phonon Induced  Attraction. $U=0$ Case}
\label{bip:ueq0}

\begin{figure}[t]
\centerline{
\includegraphics*[width=3.3in]{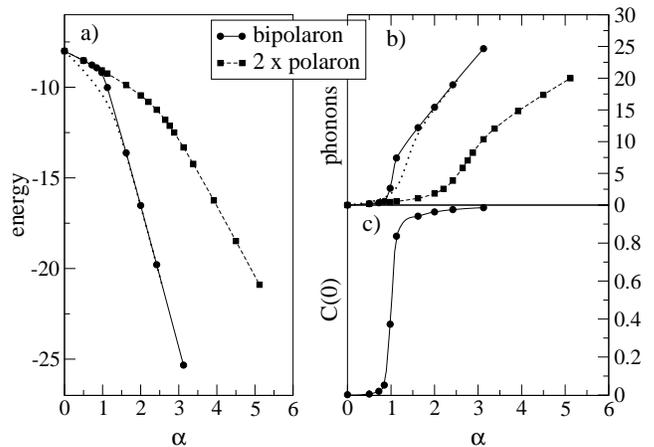}}
\caption{$U=0$, $\omega_0=1$, $t=1$. a)  The bipolaron energy  versus electron-phonon coupling (circles). 
The dashed line (squares) is $2\times$ free polaron energy. b) The bipolaron average number of phonons   
(circles) and  $2\times$ free polaron number of phonons (squares). 
c) The probability to have the
electrons on the same site, $C(0)$, in the bipolaron state. 
The dotted line in -a ( -b)  represents the energy (number of phonons) of {\it two free polarons}  versus 
the effective coupling, $\alpha_{eff}=2~\alpha$.} 
\label{fig:u0}
\end{figure}

\begin{figure}[t]
\centerline{
\includegraphics*[width=3.3in]{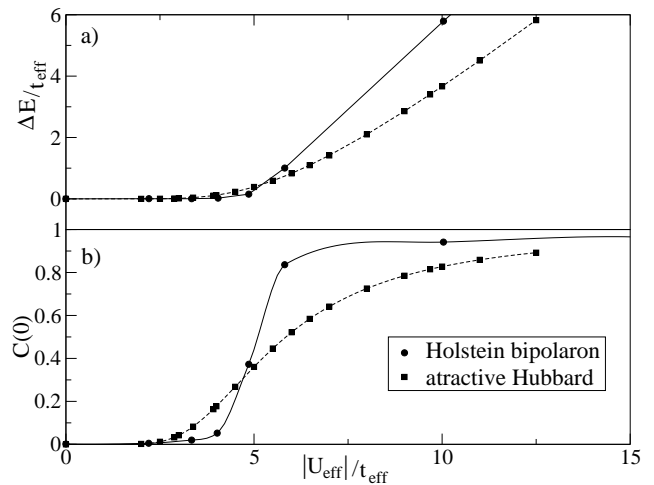}}
\caption{A comparison between HH bipolaron (solid line) and attractive Hubbard model 
(dashed line). On the horizontal axis is $U_{eff}/t_{eff}$,
where $U_{eff}=-4\alpha t$ and $t_{eff}=t_{eff}(\alpha)$ is the polaron effective hopping.
a) The binding energy in terms of $t_{eff}$. b) Probability to have two electrons on the same site.} 
\label{fig:hhcomp}
\end{figure}

With $U=0$ the effective interaction is always attractive. In 
Fig.~\ref{fig:u0} we show the evolution of the system from a very weakly bond  
bipolaron (almost two free polarons) to a strongly bound bipolaron with increasing 
$\alpha$. The transition is sharp. It can  be seen from Fig.~\ref{fig:u0}-c, where the electrons relative position
correlation function $C(r)$ (Eq.~\ref{eq:corr}) is shown at $r=0$, that in a very 
short interval around $\alpha_c = 1$ the system ground state changes from almost two free polarons to
a state where the electrons are practically on the same site. 
In comparison to the polaron case,
the critical electron-phonon coupling where the transition takes place is approximately two times smaller.
This can be understood by noticing that (see Eq.~\ref{eq:nham0}) the deformation energy at a particular site
is proportional to the square of the number of electrons on that site. Therefore for the on-site bipolaron
the effective  $\alpha$  is two times larger than the corresponding polaron one.
Thus, the bipolaron energy at a particular coupling $\alpha$ (in the strong coupling regime) is equal to
two times the free polaron energy corresponding to a double $\alpha$. The same is true for the
average number of phonons in the bipolaronic cloud (which is also proportional to the square of the number
of electrons).
This is shown with  dotted line in Fig.~\ref{fig:u0} -a and -b. From the same figures it can  also be observed
that the bipolaron transition is very sharp compared to the large to  small polaron transition~\footnote{The
Holstein polaron properties were calculated with a DQMC code similar to the one presented in~\cite{prokofev1}
and used for the Fr\"{o}hlich polaron study.}.

In Sec.~\ref{sec:perturb} we have shown that in the antiadiabatic limit 
(i.e when $\omega_0 \longrightarrow \infty$) the effective attraction induced by phonons becomes 
instantaneous and as a consequence the HH model is 
equivalent to a an attractive Hubbard model.
The attractive Hubbard Hamiltonian was considered as a realistic model to explain
the properties of systems like amorphous semiconductors~\cite{anderson,street,alder} or  
high $T_c$ superconductors~\cite{alexandrov} and was under investigation in the past~\cite{robasz}.
However, when the phonon frequency is finite, the interaction becomes retarded and
the HH physics will differ from the attractive Hubbard one.
Therefore, we think that a comparison of the  HH model
and the  attractive Hubbard model is necessary.
Aside  from inducing  a retarded  effective attraction, the other main effect of the electron-phonon interaction
is to dress  the electrons, increasing their effective mass or equivalently  reducing their 
effective hopping. 
Consequently, in Fig.~\ref{fig:hhcomp} we compare the HH Hamiltonian (with $U=0$)
with the corresponding attractive Hubbard model, defined by $U_{eff}=-4\alpha t \equiv -2 E_p$ 
(see  Eq.~\ref{eq:wcva} and  Eq.~\ref{eq:nham0})
and polaron effective hopping $t_{eff}=t_{eff}(\alpha)$. The effective polaron hopping as a function of
electron-phonon coupling was calculated numerically with a DQMC code. 
Increasing $U_{eff}/t_{eff}$ the system evolves in both cases from two free electrons state to
a state where the electrons are mainly on the same site ($S0$ bipolaron).
At small $U_{eff}/t_{eff}$, namely when $\alpha$ is smaller than the transition critical $\alpha_c$, 
the effective attraction induced by phonons is weaker than
the corresponding Hubbard attraction. The transition to  the $S0$ bipolaron
is very sharp for the HH model, unlike the attractive Hubbard model where it
is rather smooth. We found (not shown) that the smaller $\omega_0$,
the sharper the transition is. This is also in agreement with the adiabatic limit 
calculation ($\omega_0/t \longrightarrow 0$) where the transition 
is of first order~\cite{aubry}. Not only is the transition  different but
also the properties of the bipolaron at large coupling are different in the two cases.
The HH $S0$ bipolaron has a large effective mass proportional to  $t^2_{eff}$, 
thus  with a  factor of $e^{\textstyle 4\alpha / \omega_0}$ larger than the free electron mass~\cite{bonca}.
On the other hand, the effective mass of the attractive Hubbard bipolaron increases linearly with 
$|U_{eff}|=4 \alpha t$
in the large $U$ regime, as it can easily be shown analytically.

From both Fig.~\ref{fig:u0} and Fig.~\ref{fig:hhcomp} we can conclude that for the HH model there is a 
very narrow
transition region where the system evolves from  two almost free light polarons to a very heavy
$S0$ bipolaron. Before the transition, the effect of the electron-phonon
interaction is small, especially when the phonon frequency is small. Increasing the phonon frequency
results in  increasing  effective attraction.
The physics after the transition 
is well described by the strong coupling theory. Now the energy and the number of phonons is proportional 
to $\alpha$ ($E=-8 \alpha t$ and  $N_{ph}=8 \alpha t / \omega_0 $), as it can be seen in Fig.~\ref{fig:u0}-a 
and -b, and the effective mass is proportional to $e^{\textstyle 4\alpha / \omega_0}$. Unlike the weak coupling regime,
here a smaller $\omega_0$ results in a heavier bipolaron state.

\subsection{$S0$ Bipolaron to $S1$ Bipolaron Transition. $U \neq 0$ Case}
\label{bip:uneq0}

The weak coupling regime is characterized by the formation of 
a  weakly bound state. The binding energy is  extremely small even for $U=0$,
as can be seen from  Fig.~\ref{fig:u0}.  
As discussed in Sec.~\ref{sec:weakperturb}, the bipolaron binding energy  decreases
rapidly with increasing $U$, being well bellow the  resolution limit of our algorithm ($10^{-3}t$).
Therefore we are not able to determine the critical $U$ where the binding energy reaches zero.
However, we do not consider this to be a relevant problem, a bipolaron with such a small binding energy
being physically identical with a state of two free polarons.

In the strong coupling regime, 
as discussed in Sec.~\ref{sec:strongperturb}, with  increasing $U$ the system
evolves from a strongly bound $S0$ to a weakly bound $s$-wave $S1$ bipolaron. The $S0$-$S1$
transition takes place around  $U=2 E_p$.
At the critical value $U=4 E_p$ the bipolaron state ceased to exist, and the system becomes
two  polarons moving freely on the lattice. The binding energy of $S0$ bipolaron decreases linearly with 
$U$. The $S1$ bipolaron results from the exchange process and its 
binding energy is proportional to $1/U$ (Eq.~\ref{eq:Deltab}).

\begin{figure}[t]
\centerline{
\includegraphics*[width=3.3in]{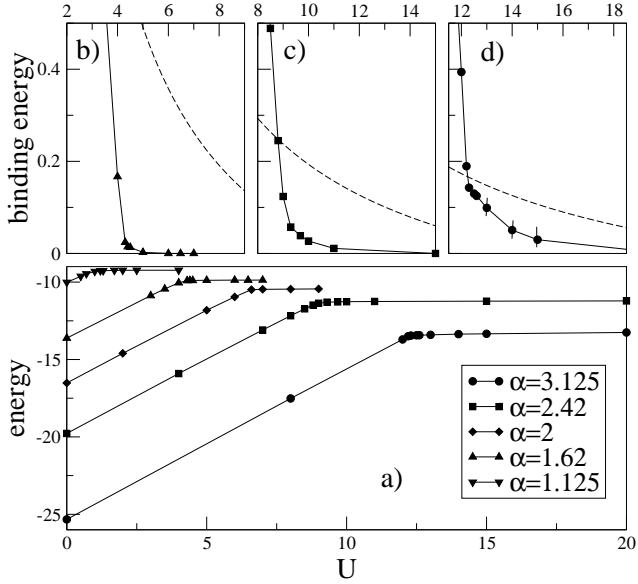}}
\caption{a) The bipolaron energy versus $U$ for different values of the electron-phonon coupling $\alpha$.
b), c) and d) The binding energy of the bipolaron in the transition region, for $\alpha=1.62$,
$\alpha=2.42$ and  respectively $\alpha=3.125$, versus $U$. The dashed line is the $S1$ strong coupling
theory predicted binding energy.} 
\label{fig:energ}
\end{figure}

\begin{figure}[t]
\centerline{
\includegraphics*[width=3.3in]{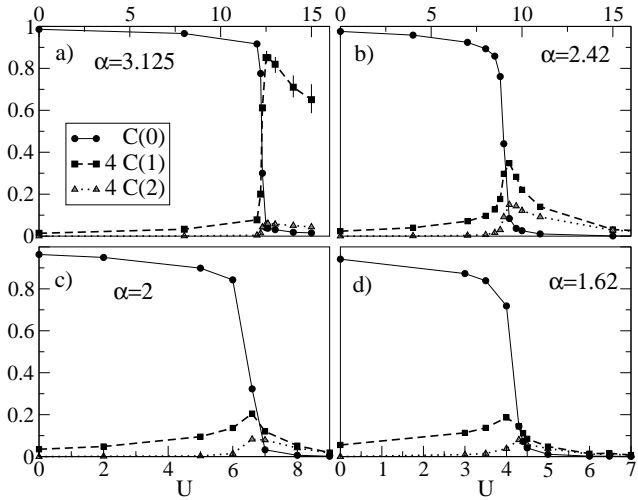}}
\caption{ The electrons position correlation function $C(0)$ , $4\times C(1)$, 
and $4\times C(2)$ (see Eq.~(\ref{eq:corr})) versus $U$
for different values of the electron-phonon coupling. } 
\label{fig:loc}
\end{figure}

\begin{figure}[t]
\centerline{
\includegraphics*[width=3.3in]{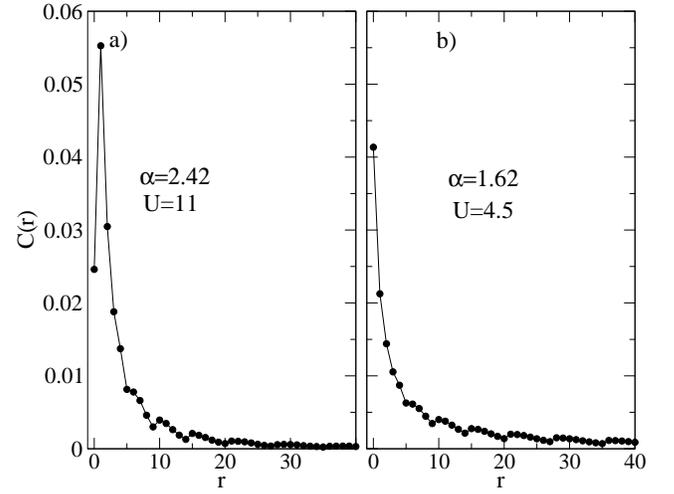}}
\caption{The correlation function $C(r)$ in the intermediate electron-phonon coupling regime.
The relative distance between electrons is given in circular coordinates.} 
\label{fig:loccor}
\end{figure}

The energy of bipolaron in the intermediate coupling regime  is presented in Fig.~\ref{fig:energ}. 
When $U=0$ the bipolaron is a $S0$ state.
With increasing $U$  the binding energy at first decreases
linearly with $U$ and afterwards its behavior changes, decreasing much slower.
The bipolaron state disappears well before  $U$ reaches  $4 E_p$.
The proportionality to $U$ of the binding energy is a characteristic of the
strongly bound $S0$ bipolaron.
From Fig.~\ref{fig:loc} it can be seen that this is correlated with  the probability to 
have the electrons on the same 
site, $C(0)$ being close to one.
At the value of $U$ where the binding energy behavior 
changes, there is a transition to a weakly bound state (with the binding energy smaller
than the one given by the strong coupling theory, Eq.~\ref{eq:Deltab}) where the probability to have the electrons on
neighboring sites is enhanced and where simultaneously  $C(0)$ drops to  small value.
For large couplings, like in  Fig.~\ref{fig:loc}-a, this state is
a small $S1$ bipolaron, with the electrons residing essentially only on the nearest-neighbor sites. For smaller
couplings, as in Fig.~\ref{fig:loc}-d, this state is a ``large'' $S1$ bipolaron, the wave function
being spread over many sites, but still with an enhanced probability that the electrons are nearest-neighbors.
This can  be seen from Fig.~\ref{fig:loccor}, where the
correlation function, $C(r)$ (Eq.~\ref{eq:corr}),
as a function of the electrons relative distance, $r$, is shown\footnote{Notice that the total probability to have
the electrons at the  distance $r$ from each other is $C(r)$ times the number of sites situated at the relative 
distance $r$. Therefore, in  Fig.~\ref{fig:loc}, we plotted  $4 \times C(1)$ and $4 \times C(2)$
which represent the probability to have the electrons on nearest and  respectively next-nearest neighbor sites.}.

\begin{figure}[t]
\centerline{
\includegraphics*[width=3.3in]{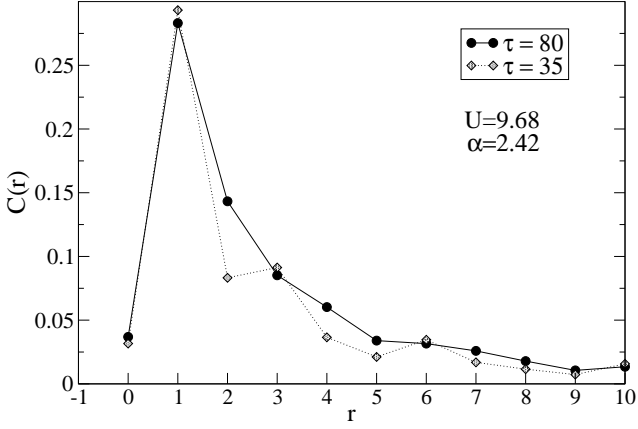}}
\caption{The correlation function $C(r)$ in the intermediate electron-phonon coupling regime for a value
of $U$ which corresponds to a large $S1$ bipolaron. The sites
at $r=2$, $r=5$ and $r=9$ are along the diagonal directions.
The statistics are collected at two different imaginary
times. For the smaller $\tau=35$ time the results represent the average of the $s$-wave symmetry
 ground state and the $d$-wave symmetry first excited state (notice the small occupation probability of 
the  sites along the diagonal directions).
At the larger time, $\tau=80$,  only the $s$-wave ground state remains.}
\label{fig:locsd}
\end{figure}

We know that, in the strong coupling regime for large $U$, a $d$-wave $S1$ bipolaron, 
quasi-degenerate in energy with the  $s$-wave $S1$ ground state exists.
When $\alpha$ is large (e.g. $\alpha=3.125$) our code projects out  both  $S1$ states at large imaginary time.
The results we obtain in this case represent the average of the corresponding two $S1$ bipolarons  properties. 
At smaller couplings the energy difference between the two $S1$ bipolarons is larger and
it is easier to project out the ground state  and thus to separate the two states. 
We show this in Fig.~\ref{fig:locsd}, where 
a comparison of the correlation function $C(r)$  measured  at two imaginary times,  $\tau=35$ and 
$\tau=80$ is made.
At time $\tau=35$ we see a smaller probability for the electrons to stay  along the diagonal directions. This is 
evidence that  our measurements capture  both the $d$-wave and the $s$-wave states. When
the measurements are taken at a larger time, the value of the correlation function at sites which correspond 
to the diagonal directions increases. 
For the presented case, $C(r)$ does not change sensible if the measurement time is
increased above $\tau=50$, thus  we can conclude that the asymptotic  regime  is reached
above this time.
The fact that at  $\tau=35$ we see a decrease of  $C(5)$, i.e. a decrease of the correlation function
at large distance along the diagonal directions, shows that the $d$-wave bipolaron in the
intermediate coupling region is a large state spread over many sites, like the  $s$-wave ground state.


\begin{figure}[t]
\centerline{
\includegraphics*[width=3.3in]{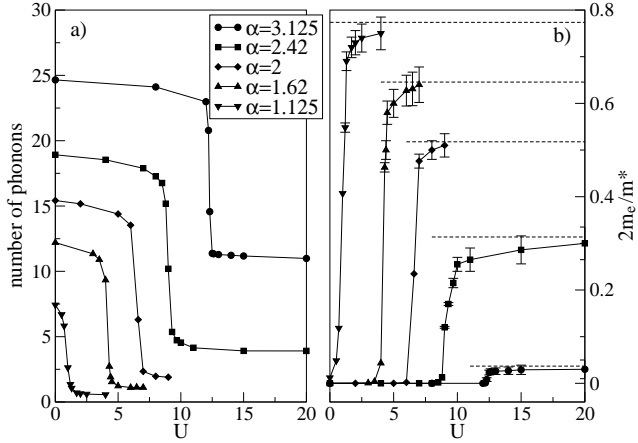}}
\caption{a) Average number of phonons in the bipolaronic cloud for different values of
electron-phonon coupling versus $U$. b) Inverse of the bipolaron effective mass versus $U$. $m_{e}$ 
is the electron mass. The horizontal lines correspond to the inverse of {\it two free polarons} effective mass.} 
\label{fig:phmass}
\end{figure}

In Fig.~\ref{fig:phmass} we address the  $S0$-$S1$ transition
by looking at the number of phonons in the  bipolaron cloud and at bipolaron effective mass.
Before the $S0$-$S1$ transition,  the number of phonons decreases  slowly with $U$, 
and it is well approximated by the strong coupling perturbation theory. After the transition
the number of phonons is roughly equal to the number of phonons corresponding to two free polarons.
The  $S0$-$S1$ transition is sharp, and the bipolaron changes from
a very heavy state ($S0$) to a light one ($S1$), as can be seen in Fig.~\ref{fig:phmass}-b where
the inverse of the bipolaron effective mass is plotted as a function of $U$. 
The $S0$-$S1$ transition is sharp even in the intermediate coupling region. 
We  also have found (not shown) that for smaller $\omega_0$ the transition is sharper. 
However, we want to specify that when we talk about  transition we mean a continuous change of
system properties and not a non-analytical jump.

To conclude, in the intermediate coupling regime, with increasing $U$, the system evolves
continuously from a heavy, strongly bound $S0$ bipolaron to a light, weakly bound state spread over many sites. 
This state,  which we call large $S1$ bipolaron, has $s$-wave symmetry and  an enhanced 
probability that the electrons occupy nearest-neighbor sites. In the same region
of parameters another stable state with $d$-wave  symmetry exists, with a smaller binding energy.
The spatial extent of this state is also large. 
When $\alpha$ is increased both $s$-wave $S1$ and $d$-wave $S1$ bipolaron wave functions become
more  localized evolving to the states predicted by the perturbation theory.
The energy difference between the $s$-wave and the $d$-wave states becomes exponentially small at large 
electron-phonon coupling. 

The results presented up to here are calculated for $K=0$ and in  the singlet channel. For the strongly
 bound $S0$ bipolaron,
in agreement with the exponentially small  effective hopping predicted by the perturbation
theory, we found a flat dispersion resulting
in a narrow  band, of the order of the calculation error bars. 
We are not able to compute the  momentum dependent  properties of the weakly bound  bipolarons
for  reasons described in Sec.~\ref{bip:dqmcdis}. 
In the triplet channel we found no bound states for any value of the parameters.

\subsection{Phase Diagram}
\label{bip:phase}
 
\begin{figure}[t]
\centerline{
\includegraphics*[width=3.3in]{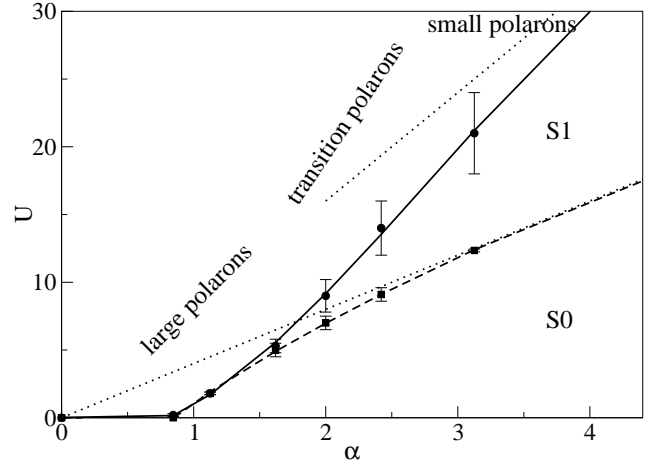}}
\caption{ Phase diagram. The solid line is the {\it bipolaron -two free polarons}  boundary. 
The dashed line separates
the $S0$ and  $S1$ bipolarons. The doted lines are the strong coupling theory results.}
\label{fig:phasebip}
\end{figure}

The phase diagram is shown in Fig.~\ref{fig:phasebip}.
We want to remind the reader that with our technique, the
calculation becomes  difficult when the binding energy is small.
The smaller is the binding energy, the larger time computations are needed.
The most difficult computations  are at both large electron-phonon  coupling and large $U$. 
The large imaginary time simulations  are difficult 
because the number of phonons and the effective mass is always large in this case (the ground state consists of two 
weakly interacting small polarons, and
a small polaron has itself an exponentially large mass and contains a large number of phonons).  
Therefore the largest errors we get are in the determination of the {\em bipolaron-two free polarons} 
boundary at large values of $\alpha$.
In the strong coupling theory this boundary is determined 
by the critical value $U=8 \alpha t \equiv 4 E_p$. In the intermediate coupling regime we find that
the bipolaron state disappears much before that value. The value  $U=8 \alpha t \equiv 4 E_p$ should be taken 
as an upper limit for the existence of the bipolaron state, reached  asymptotically when $\alpha$ is increased.

The $S1$ region contains a weakly bound state
where the probability to have the electrons on nearest-neighbor sites is larger than of having them
on the same site. Depending on the value of electron-phonon coupling
the  $S1$ bipolaron can be a large state with  the wave function spread over may sites,
or a small state where the electrons are residing on nearest-neighbor sites.
The large $S1$ bipolaron breaks into two large or  intermediate (transition) polarons
and the small  $S1$ bipolaron evolves into two small polarons with increasing  the Coulomb repulsion 
$U$. 
The $S1$ bipolaron is a state which forms only at intermediate and large electron-phonon
coupling, thus only where the polaron kinetic energy is strongly reduced. In this region,
the exchange attraction which is weakly renormalized by the phonons, can overcome the 
effective kinetic energy resulting in the formation of  the $S1$ bound state. 

At small $\alpha$, the binding energy is extremely small, well beyond the resolution limit of 
our algorithm, and decreases rapidly with increasing $U$. The maximum critical $U$ where the bipolaron
breaks can be theoretically as large as $U=4 \alpha t \equiv 2 E_p$, but for $U$ larger
than the boundary shown in Fig.~\ref{fig:phasebip} the binding energy is so small that can be safely 
approximated by zero.

\section{Conclusions}
\label{bip:conclusions}

In this paper we studied the two-dimensional HH bipolaron  using
a Diagrammatic Quantum Monte Carlo algorithm which computes the zero temperature Matsubara Green's 
functions.
The bipolaron properties are extracted from the  Green's functions behavior at large
imaginary time where the ground state is projected out. Unlike the 
other DQMC simulations used for studying  different polaron
and exciton models, in order to avoid the sign problem, our algorithm produces and sums   real space 
(Wannier orbitals  basis) diagrams.  
The code can be relative easily modified for  other  bipolaron models with 
longer range electron-phonon and/or electron-electron interaction. The dimensionality and the lattice
symmetry can also be modified.

We  calculated the phase diagram in the parameter space defined by $U$ and $\alpha$.
Depending on the parameters value, different kinds of bound states  are formed.
We studied both their properties and  the transition from one bipolaron type 
to another.

At small electron-phonon coupling two electrons form a  extremely weakly bound state when $U=0$. 
In this regime the binding energy  decreases fast with increasing the Coulomb repulsion
and  increases with increasing  phonon frequency.

For larger couplings, depending on the value of  $U$, 
the phonon-induced attraction may result in the formation of a strongly bound
$S0$  bipolaron or of a  weakly bound $S1$ bipolaron.
The $S0$ bipolaron forms at small values of $U$, for couplings larger than $\alpha_c \approx 1$.
It is an on-site state and its properties are well described by the strong coupling 
perturbation theory. With increasing $U$, around the value given by the strong coupling
perturbation theory ($U=2E_p$), the $S0$ bipolaron  transforms sharply (but continuously)
into a weakly bound state with an enhanced probability to have the electrons on nearest-neighbor
sites, called $S1$ bipolaron. The $S1$ bipolaron is a large state spread over many lattice sites 
for $\alpha$ in the intermediate regime (which corresponds to  polaron  transition region),
and  becomes nearest-neighbor localized at large $\alpha$.
The unrenormalized exchange energy  which wins over the reduced 
polaron effective hopping is responsible for binding the $S1$ bipolaron.
The binding energy of the $S1$ bipolaron
and the critical Coulomb repulsion $U$ where the bipolaron state disappears are smaller than the  values 
obtained in the strong coupling perturbation theory ($U=4E_p$).

We found that  the ground state always has $s$-wave symmetry. 
In the intermediate and strong electron-phonon coupling regime, for values of $U$ which correspond 
to the  $S1$  ground state,  an excited  $d$-wave stable state also exists.
This state is spatially large for intermediate $\alpha$ and nearest-neighbor localized for large $\alpha$,
similar to the corresponding $s$-wave  ground state. The excitation energy (difference between the
$d$-wave and $s$-wave $S1$ states) is larger 
at intermediate coupling and goes exponentially to zero when $\alpha$ is increased.

\section*{ACKNOWLEDGMENT}
We want to thank N. Prokof'ev, A.S. Mishchenko, M. Mostovoi, D. Khomskii and S. A. Trugman for 
useful discussions related to bipolarons and polarons.
We are particularly grateful the first two of them for their help in
developing the DQMC  code. 
The work was supported  by the Netherlands Foundation for Fundamental Research 
on Matter (FOM) with financial support from the Netherlands Organization for 
Scientific Research (NWO) and the Spinoza Prize Program of NWO and by  NSF 
grants DMR-0073308 and DMR-0113574.

\end{document}